\input{psfig}
 \documentclass[preprint]{aastex}

\def\lax    {${_<\atop^{\sim}}$}
\def\gax    {${_>\atop^{\sim}}$}

\def\etal   {{\it et al.}~}
\def\lya    {Ly$\alpha$~}

\shorttitle{The z=5.8 quasar}
\shortauthors{Mathur}

\begin{document}


\title{The z=5.8 Quasar SDSSp J1044-0125: A Peek at Quasar Evolution?}


\author{Smita Mathur}
\affil{Astronomy department, The Ohio State University, Columbus, OH 43210}
\email{smita@astronomy.ohio-state.edu}


\begin{abstract}
 The newly discovered z=5.8 quasar SDSSp J104433.04-012502.2 was
  recently detected in X-rays and found to be extremely X-ray
  weak. Here we present the hardness ratio analysis of the XMM-Newton
  observation. We consider various models to explain the detection in
  the soft X-ray band and non-detection in the hard band, together
  with its X-ray weakness. We show that the source may have a steep
  power-law slope, with an absorber partially covering the continuum.
  This may be X-ray evidence to support the argument of Mathur (2000)
  that narrow line Seyfert 1 galaxies, which show steep power-law
  slopes, might be the low redshift, low luminosity analogues of the
  high redshift quasars.  Heavily shrouded and steep X-ray spectrum
  quasars may indeed represent the early stages of quasar evolution
  (Mathur 2000, Fabian 1999) and SDSSp J104433.04-012502.2 is possibly
  giving us a first glimpse of the physical evolution of quasar
  properties.
\end{abstract}




\section{Introduction}

 The discovery of a quasar at redshift z=5.8 (Fan \etal 2000) was
  remarkable as it showed that luminous quasars with massive
  black-holes had formed when the Universe was less than a Gyr
  old. The quasar SDSSp J104433.04-012502.2 (SDSS 1044-0125 here
  after) was discovered as a part of the Sloan Digital Sky Survey
  imaging multicolor observations. The follow-up spectrum with Keck
  revealed a rich spectrum of a quasar at z=5.8 with strong \lya
  forest. Brandt \etal (2001, Paper I here after) then observed the
  quasar in X-rays with XMM-Newton. The quasar was clearly detected,
  but found to be unusually X-ray weak. The source was too faint to
  do meaningful spectral analysis.

 To extract spectral information from only a few tens of counts, X-ray
 astronomers have traditionally used a hardness ratio analysis
 (Maccacaro \etal 1988). The observed photons are divided into two
 energy bands, soft (S) and hard (H) for the low energy and high
 energy respectively. The hardness ratio HR=H$-$S$/$H$+$S then gives a
 measure of the shape of the X-ray spectrum. A more sophisticated
 version of the hardness ratio method has been used by various authors
 to extract valuable information on high redshift quasars: Bechtold
 \etal (1994) found that the radio-quiet high redshift quasars either
 have steeper power-law slopes and large absorbing columns or no
 absorption and similar power-law slopes than their lower redshift
 cousins. Based on a detection with {\it Einstein}, Mathur \& Elvis
 (1994) concluded that the high redshift radio-loud quasar
 GB~1508+5714 either has an unusually hard power-law slope or is
 highly absorbed. Later observations have confirmed the unusually hard
 slopes of high redshift radio-loud quasars (e.g. Moran \& Helfand
 1997, Reeves \etal 2001). The hardness ratio analysis thus provides
 us with a powerful tool to extract spectral information when the
 total observed counts are too small to perform detailed spectral
 analysis.

 In this paper we perform the hardness ratio analysis of the highest
 redshift quasar SDSS 1044-0125 to expand upon the work
 already reported in Paper I.

\section{XMM Observations of SDSS 1044-0125}

 SDSS 1044-0125 was detected with XMM EPIC-pn. The total counts
 observed in the 0.5--2.0 keV band were 32$\pm$9. This corresponds to
 the rest frame energy range of 3.4--13.6 keV (see Paper I for the details of observation). Based on this observed
 count rate, SDSS 1044-0125 is unusually X-ray weak; it has about a
 factor of ten smaller X-ray flux for its optical observed flux
 relative to other luminous optically selected quasars (Paper
 I). These observations imply that either (1) the object is
 intrinsically X-ray weak, and so a very unusual quasar, or (2) it is
 a normal quasar that is heavily absorbed with a column density
 N$_H$\gax $10^{24}$ cm$^{-2}$, as seen in some Broad Absorption Line
 Quasars (BALQSOs).  Brandt \etal have discussed both of these
 possibilities, though the latter was their preferred interpretation.

 Note also that SDSS 1044-0125 was not detected in the 2.0--7.0 keV
 band (rest frame 13.6--47.6 keV). Since hard X-rays are less affected by
 photoelectric absorption, the non-detection in the hard band was
 rather surprising if heavy absorption was indeed the cause of X-ray
 weakness of the object\footnote{We note that Brandt \etal have
 mentioned that perhaps partial covering of the X-ray continuum source
 may explain the photons detected at \lax 8 keV}. In most BALQSOs, for
 example, X-ray detections are preferentially in the hard band (Mathur
 \etal 2000 and references therein).  Another point to note
 is that the low spectral resolution X-ray observations cannot determine
 the redshift of the absorber. Absorption at lower redshift with smaller
 column density may mimic the absorption at high redshift with larger
 column density of gas. Given that SDSS 1044-0125 is at very high
 redshift, and that there is a lot of absorption along the line of
 sight including that due to a Lyman limit system and a MgII system (Fan
 \etal 2000), the X-ray absorption may take place in the intervening
 systems.  To study these different possibilities and
  to extract  further spectral information, we perform
 a hardness ratio analysis of SDSS 1044-0125.

\subsection{The hardness ratio analysis.}

 For SDSS 1044-0125, the total counts detected in the soft band are
31.7$\pm 8.5$ and the upper limit in the hard band is 12.3 counts
(95\% confidence, Paper I). This results in the hardness ratio HR$<
-0.44$. The maximum HR (assuming minimum soft counts, $-1\sigma$) is
$-0.31$. In figure 1 we have plotted the expected HR as a function of
X-ray power-law slope $\Gamma$ (1+$\alpha$ for f$_{\nu} \propto
\nu^{-\alpha}$) for a range of absorbing column densities. We have
used the 0.5--2.0 keV range for the soft band and 2.0--7.0 keV range
for the hard band as in Brandt \etal. All the HR calculations were
performed using the Portable, Interactive, Multi-Mission Simulator
(PIMMS) version 3.0 (Mukai 2000) and XSPEC (Arnaud 1996) was used to
define the spectral models. (See Paper I for the appropriateness of
using PIMMS).  The lowest curve corresponds to Galactic column density
of 4.6$\times 10^{20}$ cm$^{-2}$ towards SDSS 1044-0125, next curve to
column density ten times Galactic, and the top one to column density
of 1.7$\times 10^{22}$ cm$^{-2}$. These are all column densities at
z=0.0. The highest column density plotted in Figure 1 is that required
for the suppression of the soft flux by a factor of ten and would
correspond to \gax $10^{24}$ cm$^{-2}$ at z=5.8. If the quasar is not
intrinsically X-ray faint, then this much absorption is required to
suppress the total flux to the observed value. What is immediately
apparent from Figure 1 is that the required excess absorption is
inconsistent with the observed hardness ratio {\it for any reasonable
value of $\Gamma$}.

 What is, then, the possible spectral shape of SDSS 1044-0125
 consistent with the observed hardness ratio? A simple power-law with
 just Galactic absorption and slope in the normal range $\Gamma
 \approx 2$, is consistent with observations (also noted by Brandt
 \etal). This, however, makes the quasar unusually X-ray weak. There
 are a few ways to make the heavy absorption scenario consistent with
 the observed hardness ratio: (1) a high energy cutoff in the observed
 band; this will bring down the number of hard X-ray photons, (2) a
 completely Compton thick X-ray absorber suppressing both soft and
 hard X-ray flux, but partial covering of the continuum source leaking
 out some soft photons, and (3) a partially covering absorber and a 
 steep X-ray spectrum. Below we consider each of these possibilities
 in turn. One immediate conclusion of the above exercise is that the
 absorber must be close to the continuum source for partial covering
 to work. So, it is highly unlikely that the X-ray absorption is
 caused by the intervening matter.

\subsubsection{High energy cutoff.}

 High energy cutoffs to the power-law spectra are expected from the
 thermal Comptonization models and are required to explain the cosmic
 hard X-ray background. There is no universal value for a cutoff
 energy, but generally it is assumed to be $\approx 270$ keV (e.g Matt
 1998). This is too high to have any effect on the observed hard band
 (rest frame 13.6--47.6 keV). To our knowledge, the lowest value of
 high energy cutoff observed so far is in the highly absorbed maser
 source ESO 103-G35 (Wilkes \etal 2000). It may have a cutoff
 energy as low as $30\pm 10$ keV. We calculated hardness ratios for
 cutoff energies of 20, 30 and 40 keV for $\Gamma=2.0$ and the
 e-folding energy of 20 keV. The resulting HR values are +0.22, +0.31,
 and +0.34 respectively. There values are well above the observed HR
 for SDSS 1044-0125. We then repeated the above experiment for steeper
 power-law slopes. The results indicate that if the high energy cutoff
 model is to account for the observed HR, then the cutoff energy has
 to be smaller than 20 keV {\it and} $\Gamma$ has to be steeper than
 3. This, again, is extremely unusual.

\subsubsection{Compton thick absorber with partial covering.}

 If an absorber near the source is completely Compton thick (say,
 N$_H=10^{25}$ cm$^{-2}$), then the continuum is completely suppressed
 even in the high energy band. This happens because the high energy
 photons are down-scattered to lower energy where they are effectively
 absorbed by photoelectric absorption (Matt 1998). If such an
 absorber does not cover the source completely, with 10\% of the
 continuum leaking through, then the observed X-ray flux is naturally
 $1/10$ of the intrinsic flux. This can happen either as a result of
 the geometry of the absorber or as a result of scattering towards our
 line of sight. Such a scenario with a normal power-law slope of
 $\Gamma=2$ then explains the observations including the hardness ratio
 ($\S 2.1$). We have verified with explicit simulations that indeed,
 for a model with $\Gamma=2$, and an absorber with 90\% covering, a
 column density of as much as N$_H=10^{25}$ cm$^{-2}$ is required to
 suppress the total flux by a factor of ten.

 Such a model, however, has other consequences. The absorber in this
 case is not only thick to Compton scattering, it is also thick to
 Thomson scattering. Thomson scattering is achromatic (up to $m_e
 c^2$) and would result in attenuation at all wavelengths including
 optical and UV (unless the absorber covers the whole sky as seen by
 the source). For the column density quoted above,
 $\tau_{Thomson}=6.65$ with attenuation factor of 773. Thus the source
 would be heavily attenuated not only in the X-rays, but also in
 UV. This is not unusual. Broad absorption line quasars, which show
 strong X-ray absorption, also suffer from attenuation in the
 optical/UV, by an amount appropriate for their observed column
 densities (e.g. Mathur \etal 2000 and references there in). In NGC
 1068, a classic example of Seyfert 2 galaxies, in which the X-ray
 absorber is believed to be Compton thick (e.g. Guainazzi, Matt \&
 Fiore 1999), the optical and UV continuum is also completely
 obscured. 

 On the other hand, SDSS 1044-0125 is luminous in optical (rest frame
 UV). As discussed in $\S 1$, even at the observed flux levels, the
 luminosity of the object requires a black hole (BH) of $3\times10^9$
 M$_{\odot}$ accreting at the Eddington rate. If the intrinsic
 luminosity of the source were significantly higher, then it would
 push the required BH mass to significantly higher values. In the
 framework of hierarchical models of structure formation, Haiman \&
 Loeb (2001) have calculated the maximum allowed BH mass as a function
 of redshift that could be found in a given survey. At z=5.8 and BH
 mass of $3\times 10^9$ M$_{\odot}$, SDSS 1044-0125 is already at a
 model limit (see their figure 3). While not impossible, orders of
 magnitude increase in the implied BH mass would be stretching the
 limits. Moreover, if the observed optical flux of SDSS 1044-0125 is
 also significantly attenuated, then it does not solve the problem of
 X-ray weakness of the object.

 One way out of this problem is to place the Compton thick X-ray
 absorber outside the X-ray emitting region, but inside the region
 emitting UV. This would be extremely difficult given that the
 X-ray/UV emission arises from a few to a few tens of Schwarzschild
 radii away from the nuclear black hole, and especially if UV photons
 provide the seed population for Compton up-scattering to produce
 X-rays (e.g. Nandra 2001).

\subsubsection{A steep X-ray spectrum and a partially covering absorber}

To alleviate the above problem, we considered a model in which the
X-ray continuum of the source has a steep spectrum and, again, a
partially covering absorber along our line of sight. A steeper
continuum would reduce the hard X-ray flux for a given normalization,
while the absorber would suppress the soft X-ray flux. The required
column density of the absorber then may not be excessive. Such a model
is physically motivated. SDSS 1044-0125 seems to be a broad absorption
line quasar (Maiolino \etal 2001). There is some evidence that the
X-ray power-law slopes of BALQSOs may be steep (Mathur \etal 2001;
although see Green \etal 2001).

So we constructed a model with $\Gamma=2.5$ and the covering factor of
the absorber 90\%. We then varied the column density of the absorber
from N$_H=10^{22}$ to 10$^{25}$ cm$^{-2}$. In each case we noted the
factor by which the intrinsic flux was suppressed and calculated the
hardness ratio. A factor of ten smaller flux (compared to an
unobscured source with $\Gamma=2$) is obtained for N$_H\sim 7 \times
10^{23}$ cm$^{-2}$. The hardness ratio in this case is between $-0.47$
and $-0.58$, consistent with the observed value. Thus such a model
satisfies constraints from both flux and HR. The implied Thomson
opacity in this case is $\approx 0.5$, resulting in no significant
attenuation in optical/UV. The derived column density is also
consistent with that observed in other BALQSOs (Mathur \etal 2000,
2001, Green \etal 2001, Gallagher \etal 2001).

\section{Discussion}

 We are aware that only 33 counts were detected in the XMM observation
 of SDSS 1044-0125. Trying to extract much more information from this
 observation might seem like over-interpreting the data. However, the 
 hardness ratio analysis has been used in situations like this. This
 paper not only shows the power of this technique, but also draws
 important conclusions from its use.

 We show that the X-ray properties of this z=5.8 quasar are unlikely
 to be to exactly like its low-redshift cousins. It is either
 extremely X-ray weak (Paper I) or it has a steep power-law
 slope. Since we are looking at an object at the time when the
 structures in the universe were very young, it may be even more of a
 surprise if it looked exactly like nearby quasars. Quasars must
 evolve, and the evolution of quasar luminosity function has been
 observed. However, we still do not know $\it how$ quasars
 evolve. With the observation of SDSS 1044-0125 we might be getting a
 first look at the evolution of quasar spectra with time. Heavily
 shrouded and steep hard X-ray spectrum quasars may represent an early
 evolutionary stage (Fabian 1999, Mathur 2000). Given the extremely
 large luminosity of the object (Fan \etal 2000), it is most likely
 accreting at an Eddington, or even super-Eddington rate. The narrow
 line Seyfert 1 galaxies (NLS1s), nearby AGNs believed to be accreting
 at close to Eddington rate, also have steep hard X-ray spectra
 (Brandt, Mathur \& Elvis 1997). Mathur (2000) has argued that the
 NLS1 galaxies might be low redshift, low luminosity analogues of the
 high redshift quasars. Here we might be seeing X-ray evidence in
 support of that argument. But can accretion onto a $10^9$ M$_{\odot}$
 black hole produce X-rays with such steep power-law slopes? It would
 challenge current accretion disk theories. Clearly, good X-ray
 spectra of radio-quiet high redshift quasars are needed to confirm
 this hypothesis.

\acknowledgments

It is my pleasure to thank David Weinberg, Niel Brandt \& Giorgio Matt
for useful discussions, Koji Mukai for updating PIMMS version 3.0, and
Giorgio Matt for the use of his Monte Carlo code. This work is
supported in part by NASA grant NAG5-8913 (LTSA).



%% new page.

\clearpage

%% No more than seven \figcaption commands are allowed per page,
%% so if you have more than seven captions, insert a \clearpage

%% There must be a \figcaption command for each legend. Key the text of the
%% legend and the optional \label in curly braces. If you wish, you may
%% include the name of the corresponding figure file in square brackets.
%% The label is for identification purposes only. It will not insert the
%% figures themselves into the document.
%% If you want to include your art in the paper, use \plotone.
%% Refer to the on-line documentation for details.

\begin{figure}[h]
\psfig{file=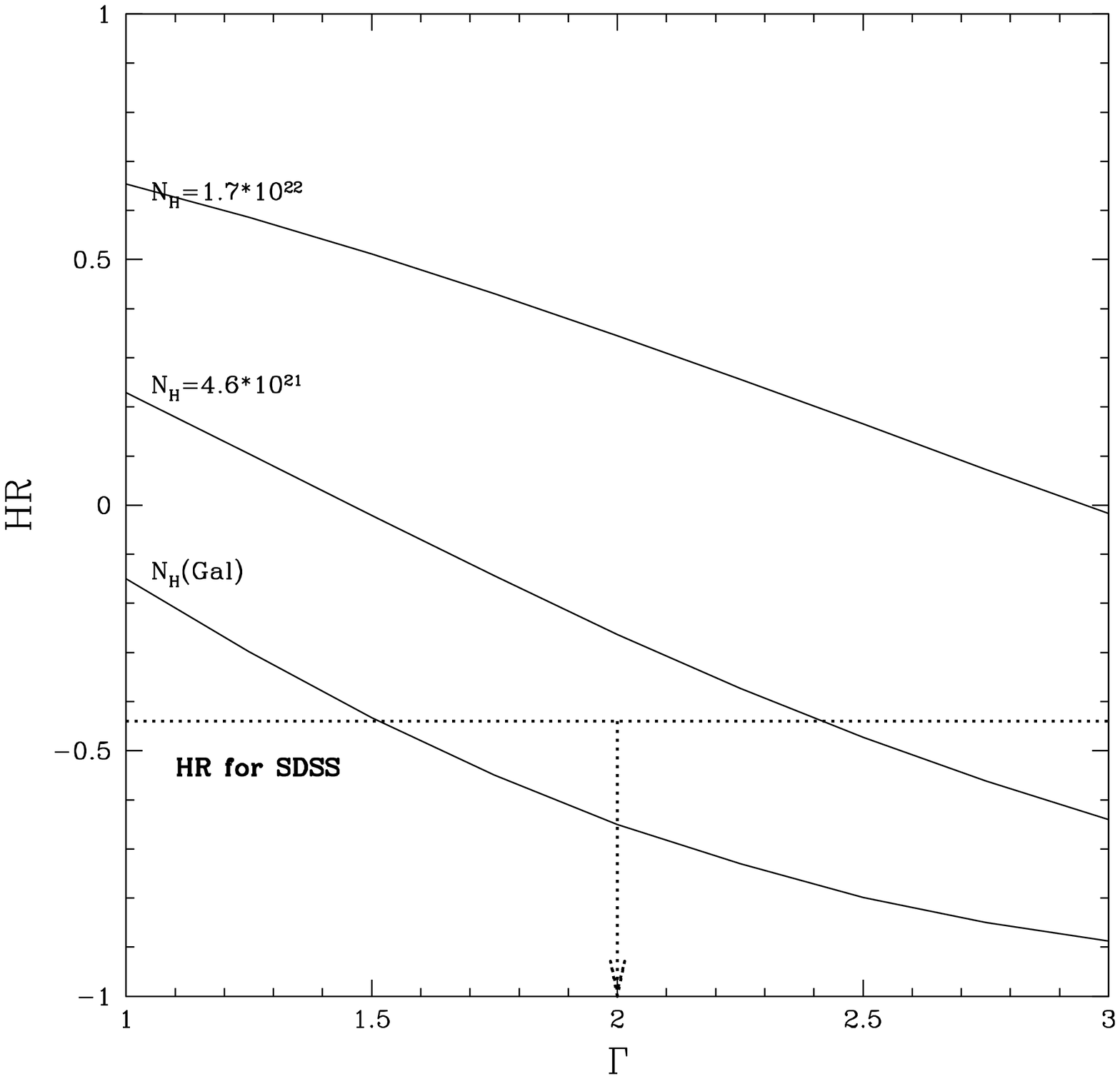,height=5in,width=5in}
\caption{Calculated hardness ratio is plotted as a function of
X-ray power-law photon index for a range of absorbing column
densities at z=0. The horizontal line represents the observed upper limit for
SDSS 1044-0125. The top curve corresponds to the column density 
required to suppress the soft X-ray flux by a factor of 10. This model
is inconsistent with observed HR for any reasonable value
$\Gamma$. The observed HR is consistent with only Galactic absorption,
but that will make the quasar extremely X-ray weak. \label{fig1}}
\end{figure}

%% Tables should be submitted one per page, so put a \clearpage before

%% Two options are available to the author for producing tables:  the
%% deluxetable environment provided by the AASTeX package or the LaTeX
%% table environment.  Use of deluxetable is preferred.
%%

%% Three table samples follow, two marked up in the deluxetable environment,
%% one marked up as a LaTeX table.

%% In this first example, note that the \tabletypesize{}
%% command has been used to reduce the font size of the table.
%% Note also that the \label command needs to be placed 

\end{document}